\documentclass[12pt]{iopart}
\usepackage{graphicx}

\begin{document}
\title[Particle dependence of elliptic flow]{
Particle dependence of elliptic flow in Au~+~Au collisions at
$\sqrt{s_{_{NN}}} = 200$~GeV }
\author{Paul Sorensen\dag\ for the STAR collaboration\footnote[2]{For a complete collaboration list see~\cite{Caines}}}
\address{\dag\ University of California, Los Angeles, California 90095}
\ead{sorensen@physics.ucla.edu}

\begin{abstract}
The elliptic flow parameter ($v_2$) for $K_S^0$ and
$\Lambda+\overline{\Lambda}$ has been measured at mid-rapidity in
Au + Au collisions at $\sqrt{s_{_{NN}}}=200$~GeV by the STAR
collaboration. The $v_2$ values for both $K_S^{0}$ and
$\Lambda+\overline{\Lambda}$ saturate at moderate $p_T$, deviating
from the hydrodynamic behavior observed in the lower $p_T$ region.
The saturated $v_2$ values and the $p_T$ scales where the
deviation begins are particle dependent. The particle-type
dependence of $v_2$ shows features expected from the hadronization
of a partonic ellipsoid by coalescence of co-moving quarks. These
results will be discussed in relation to the nuclear modification
factor ($R_{CP}$) which has also been measured for $K_S^0$ and
$\Lambda+\overline{\Lambda}$ by the STAR collaboration.
\end{abstract}
\submitto{\JPG} 



\section{Introduction}

The elliptic component of the event-wise azimuthal anisotropy of
particle production (i.e. {\it elliptic flow} or $v_2$) is thought
to probe the early stages of relativistic heavy ion
collisions~\cite{hydroOllitrault92}. Measurements of $v_{2}$ for
identified particles~\cite{Stv2pid,v0v2130} and charged
hadrons~\cite{aihong,Chgv2} at the Relativistic Heavy Ion Collider
(RHIC) indicate a conversion of spatial anisotropy to momentum
anisotropy near the hydrodynamical
limit~\cite{hydroOllitrault92,hydroPasi01,hydroShuryak01}. For
$p_T$ greater than 2 GeV/c, however, the charged hadron $v_{2}$
deviates from hydrodynamical calculations and saturates at a large
value approximately independent of $p_T$ up to
6~GeV/c~\cite{Chgv2}.
The measurements of $v_2$ for $K_S^0$ and $\Lambda +
\overline{\Lambda}$ also indicate a saturation at moderate $p_T$.
Models using large parton energy loss~\cite{dEdx,SurfShuryak} and
transport opacity~\cite{TransportMolnar} have been discussed in
relation to the saturation and centrality dependence of charged
hadron $v_2$ at large $p_T$.  A saturated $v_2$ could also arise
if the extent of the $p_T$ region where {\it soft processes}
dominate the spectrum is particle
dependent~\cite{MesonBaryonGyulassy}. In this case the hyperon
$v_2$ may continue to rise at intermediate $p_T$ (perhaps
following hydrodynamic model calculations) while $v_2$ of the less
massive meson decreases.  It has also been suggested that if a
partonic state exists prior to hadronization, the process of
particle formation at moderately high $p_T$, by string
fragmentation, parton fragmentation~\cite{zlin01} or quark
coalescence~\cite{CoalVoloshinv2,CoalLinv2,CoalMullerRaa,CoalKoRatios},
may lead to a dependence of $v_2$ and $R_{AA}$ on particle type.
As such, it's possible that these measurements will provide
information on the existence and nature of an early partonic
state.

In this Letter, we report the measurement of $v_{2}$ at
mid-rapidity, $|y| \leq 1.0$, for $K_{S}^{0}$ and lambda
($\Lambda$) $+$ antilambda $(\overline{\Lambda}$) with $0.4 < p_T
< 6.0$~GeV/c and $0.6 < p_T < 6.0$~GeV/c respectively in Au+Au
collisions at $\sqrt{s_{_{NN}}} = 200$~GeV.

\section{Analysis and Results}

This analysis uses $1.6\times 10^{6}$ minimum--bias trigger events
and $1.5\times 10^{6}$ central trigger events detected in the STAR
detector system~\cite{STAR}. The particles, $K_{S}^{0}$, $\Lambda$
and $\overline{\Lambda }$, were identified from the charged
daughter tracks produced in the decays $K_{S}^{0}\rightarrow \pi
^{+}+\pi ^{-}$, $\Lambda \rightarrow p+\pi ^{-}$ and
$\overline{\Lambda } \rightarrow \overline{p}+\pi ^{+}$. A
detailed description of the analysis, such as track finding, decay
vertex topology cuts, and the estimation of detection efficiency,
is given in Refs.~\cite{v0v2130,LLbar130}.

We use the yield as a function of $(\phi_{ij} - \Psi^{R}_j)$ to
calculate $v_2 = \langle \cos[2(\phi_{ij} - \Psi^{R}_j)]\rangle$,
where $\phi_{ij}$ is the azimuthal emission angle of particle $i$
in event $j$ and $\Psi^{R}_j$ is the reaction plane angle for
event $j$, where, to remove autocorrelations, the decay daughter
tracks associated with particle $i$ are excluded from its
calculation. Within statistical errors, the $\Lambda$ $v_2$ is the
same as $\overline{\Lambda}$ $v_2$, so they are summed together.

\begin{table}[hbt]
\caption{The systematic errors for minimum bias $v_2$ from
non-flow effects (n-f) and background contamination (bkg). The
values represent the {\em absolute} errors.  The $p_T$ resolution,
$\protect\delta p_T/p_T$ is also listed. } \label{syserr} \small{
\begin{tabular}{l|ccc|ccc}
\hline
~ & ~ & $K_S^0$ & ~ & ~ & $\Lambda + \overline{\Lambda}$ \\
\hline
$p_T$ (GeV/c) & 1.0 & 2.5 & 4.0 & 1.0 & 2.5 & 4.0 \\
\hline $v_2$ (bkg) & \scriptsize{+0.000 -0.001} & \scriptsize{+0.001 -0.007} & \scriptsize{+0.003 -0.018} & \scriptsize{+0.001 -0.007} & \scriptsize{+0.005 -0.001} & \scriptsize{+0.005 -0.001} \\
$v_2$ (n-f) & \scriptsize{+0.00 -0.01} & \scriptsize{+0.00 -0.04} & \scriptsize{+0.00 -0.03} & \scriptsize{+0.00 -0.01} & \scriptsize{+0.00 -0.04} & \scriptsize{+0.00 -0.04} \\

\hline
$\delta p_T/p_T$ & $0.016$ & $0.027$ & $0.037$ & $0.016$ & $0.027$ & $0.037$ \\
\hline
\end{tabular}
}
\end{table}

Possible sources of systematic error in the calculation of $v_{2}$
are correlations unrelated to the reaction plane (non-flow
effects), uncertainties in the extraction of yields from the
invariant mass distributions, the particle momentum resolution
($\delta p_T/p_T$), and biases introduced by the cuts used in the
analysis.  Table~\ref{syserr} lists the dominant systematic errors
for three transverse momenta.
The non-flow systematic error is dominant.
The non-flow effects for charged particle $v_2$ are discussed in
Refs.~\cite{Chgv2,aihong} but, the particle dependence of these
effects has not been measured. We assume a similar magnitude of
non-flow contribution to $\Lambda + \overline{\Lambda}$ and
$K^0_S$ $v_2$. A 4-particle cumulant analysis of $\Lambda +
\overline{\Lambda}$ and $K^0_S$ $v_2$ will be less sensitive to
non-flow effects but, to be conclusive, will require a larger data
sample than is currently available.


\begin{figure}[hbtp]
\centering\mbox{
\includegraphics[width=0.99\textwidth]{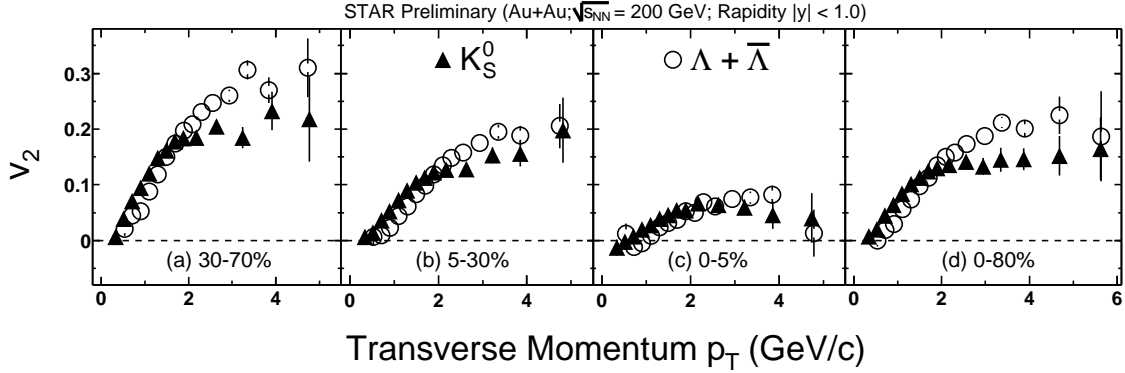}}
\caption{ The elliptic flow of $K_{S}^{0}$ and $\Lambda +
\overline{\Lambda}$ as a function of $p_T$ for 30--70\% (a),
5--30\% (b), 0--5\% (c) and 0--80\% (d) of the collision cross
section. The error bars shown are statistical errors only. }
\label{fig1}
\end{figure}


Fig.~\ref{fig1} shows $v_{2}$ of $K_{S}^{0}$ and $\Lambda +
\overline{\Lambda}$ as a function of $p_T$ for the centrality
intervals, 30--70\% (a), 5--30\% (b), 0--5\% (c), and 0--80\% (d)
of the geometrical cross section.
The $p_T$ dependence of the $v_{2}$ for all the centrality bins
has a similar trend.
There is a saturation and particle dependence at moderate $p_T$
for each of the centrality intervals and, as such, the saturation
for the minimum-bias $v_2$ (0--80\%) cannot be due to the
superposition of drastically different $p_T$ dependencies in
various centrality bins. This measurement establishes the
saturation and particle dependence of $v_2$ at the moderate to
high $p_T$ region.

\begin{figure}[hbtp]
\centering\mbox{
\includegraphics[width=0.65\textwidth]{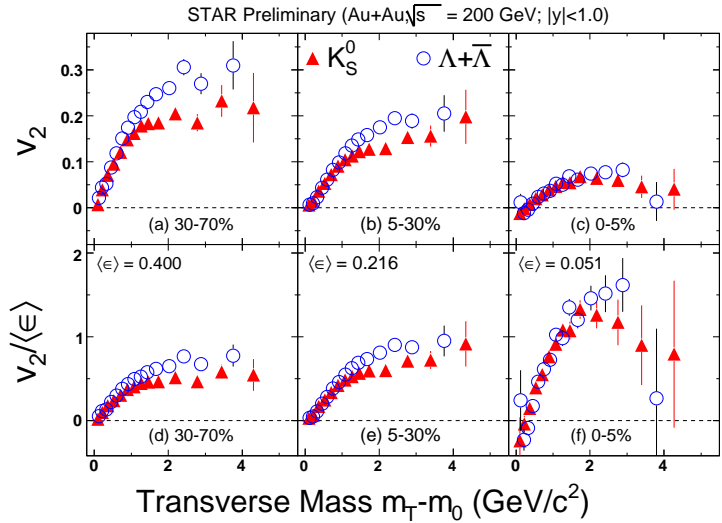}}
\caption{ (Top panel) The elliptic flow of $K_{S}^{0}$ and $\Lambda +
\overline{\Lambda}$ as a function of $m_T-m_0$ for 30--70\% (a),
5--30\% (b) and 0--5\% (c) of the collision cross
section. The error bars shown are statistical errors only. (bottom panel) The same scaled by the initial eccentricity.}
\label{fig1b}
\end{figure}

\begin{figure}[hbtp]
\centering\mbox{
\includegraphics[width=0.65\textwidth]{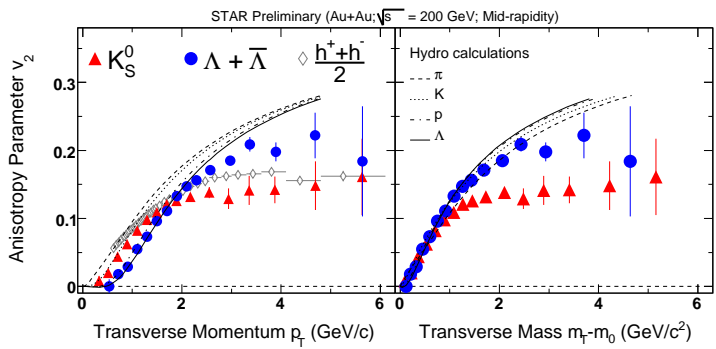}}
\caption{ The elliptic flow of $K_{S}^{0}$ and $\Lambda +
  \overline{\Lambda}$ as a function of $p_T$ (top panel) and $m_T-m_0$
  (bottom panel) for 0--80\% of the collision cross section. The error
  bars shown are statistical errors only. }
\label{fig1c}
\end{figure}

Fig.~\ref{fig1b} and Fig.~\ref{fig1c} show the same data as
Fig.~\ref{fig1} but plotted versus $m_T-m_0$ which is approximately
the kinetic energy of the hadron. In the low momentum region the
$K_S^0$ and $\Lambda + \overline{\Lambda}$ $v_2$ appears to fall on a
single straight line. The hydrodynamic calculations in
Fig.~\ref{fig1c} seem to capture this trend. In the bottom panels of
Fig.~\ref{fig1b} we've scaled the $v_2$ values by the eccentricity of
the overlap region for the various centralities. Hydrodynamic models
predict that $v_2$ should scale with this initial eccentricity.

\begin{figure}[tbph]
\centering\mbox{
\includegraphics[width=0.6\textwidth]{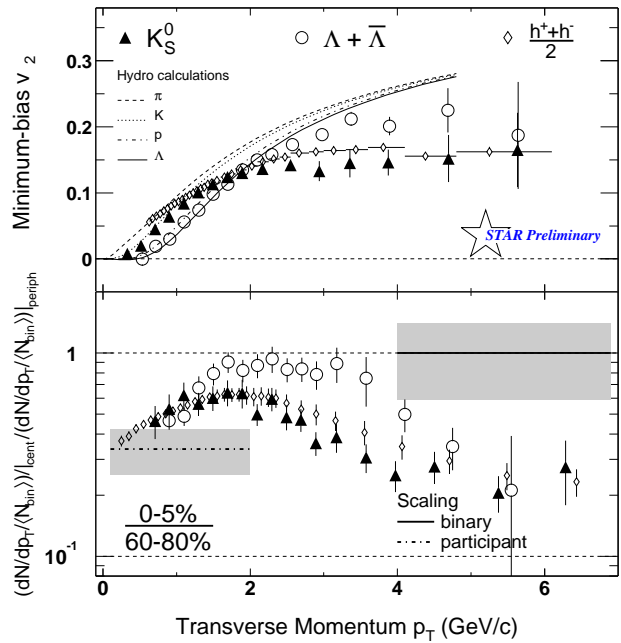}}
\caption{ Minimum-bias $v_2$ (top panel) and $R_{CP}$ (bottom
panel) for $K_{S}^{0}$ and $\Lambda +\overline{\Lambda}$. The
$v_2$ errors are statistical only. The widths of the gray bands
represent uncertainties in the model calculation of $N_{binary}$
and $N_{part}$. Charged hadron $v_2$ and $R_{CP}$ are also shown.}
\label{fig2}
\end{figure}

In Fig.~\ref{fig2} (top) we show $v_{2}(p_T)$ for $K_{S}^{0}$,
$\Lambda + \overline{\Lambda}$, and charged
hadrons~\cite{FilimonovQM2002} along with hydrodynamic model
calculations of $v_2$ for identified particles~\cite{hydroPasi01}.
Below $p_T \sim 1.2$~GeV/c $v_2$ is consistent with the
calculations and in agreement with the previous results for
$K_{S}^{0}$ and $\Lambda + \overline{\Lambda}$ at
$\sqrt{s_{_{NN}}} = 130$~GeV~\cite{v0v2130}.
Contrary, however, to hydrodynamical calculations, where at a
given $p_T$ heavier particles will have smaller $v_{2}$ values,
the measured $v_2$ of the heavier hyperon saturates at a value
significantly {\em larger} than the $v_2$ of the lighter
$K_{S}^{0}$ meson.
The $p_T$ scale where the measured $v_{2}$ deviates from the
hydrodynamical prediction is particle dependent with the hyperon
$v_2$ following the prediction up to $p_T \sim 2.0$~GeV/c while
the $K_{S}^{0}$ $v_2$ deviates much sooner.


The ratio ($R_{CP}$) of the yields in central and peripheral
collisions scaled by the number of binary nucleon-nucleon
collisions ($N_{binary}$), may also be sensitive to the effects of
energy loss and hadronization via parton
coalescence~\cite{CoalMullerRaa,dEdx}.
In Fig.~\ref{fig2} (bottom) we show $R_{CP}$ for $K_{S}^{0}$,
$\Lambda + \overline{\Lambda}$ and charged hadrons using the
centrality intervals 0--5\% (central) and $60$-$80\%$
(peripheral)~\cite{LongSQM2003,highpt200}.
The charged hadron spectrum at $p_T>2$ GeV/c for the 60--80\%
centrality bin approximately follows binary collision scaling
without medium modification~\cite{highpt200}. As such, when this
bin is used, $R_{CP}$ approximates $R_{AA}$.
The bands in Fig.~\ref{fig2} represent the expected values of
$R_{CP}$ for $N_{binary}$ and $N_{part}$ scaling including
systematic variations within the calculation~\cite{highpt200}.

For $p_T < 5$~GeV/c, the $K_{S}^{0}$ and $\Lambda +
\overline{\Lambda}$ yields are suppressed (relative to
$N_{binary}$ scaling) by different magnitudes.
In addition, the {\em $p_T$ scale} associated with the onset
suppression has a dependence on particle-type that is similar to
the dependence in $v_2$ for the onset of saturation.
At $p_T \sim 5.0$~GeV/c, $R_{CP}$ values for $K_{S}^{0}$ and
$\Lambda + \overline{\Lambda}$ are both approaching the value of
the charged hadron $R_{CP}$.

\section{Discussion}
Although $R_{CP}$ depends only on the yield in the central and
peripheral bins, and the $v_2$ in Fig.~\ref{fig2} is from a
minimum-bias centrality interval, the two parameters may be
intimately related.  The differential elliptic flow ($v_2(p_T)$)
measures the ratio of the $p_T$ spectrum of particles emitted in
the direction of the reaction plane (in-plane) to that of
particles emitted perpendicular the reaction plane (out-of-plane).
In hydrodynamic models, we expect the pressure gradient to be
larger in the in-plane direction than the out-of-plane direction.
In this case, $v_2$ will be the ratio of a $p_T$ spectrum with
more hydrodynamic flow to one with less flow. As such, taking the
ratio ($R_{CP}$) of a $p_T$ spectrum that exhibits large flow
(central) and one exhibiting less flow (peripheral) {\em should}
lead to a very similar $p_T$ and particle-type dependence. A
detailed study of $v_2$ and $R_{CP}$ for identified particles
should reveal the extent (in $p_T$ and centrality) to which
hydrodynamical models are valid.

A {\it surface emission} scenario, where partons traversing a
dense medium experience large energy losses, has been discussed in
relation to the large, $p_T$ independent $v_2$ measured for
charged hadrons~\cite{SurfShuryak}. This mechanism should also
lead to a suppression of particle production in central Au+Au
collisions.
This scenario, however, is inconsistent with STAR measurements of
$v_2$ and $R_{CP}$ for $K_S^0$ and $\Lambda + \overline{\Lambda}$.
The smaller suppression manifested in the $\Lambda +
\overline{\Lambda}$ $R_{CP}$ contradicts the larger azimuthal
anisotropies manifested in $\Lambda+\overline{\Lambda}$ $v_2$
values. In addition, calculations based on a surface emission
model~\cite{SurfShuryak} cannot produce $v_2$ values as large as
those measured.

\begin{figure}[hbtp]
\centering\mbox{
\includegraphics[width=0.8\textwidth]{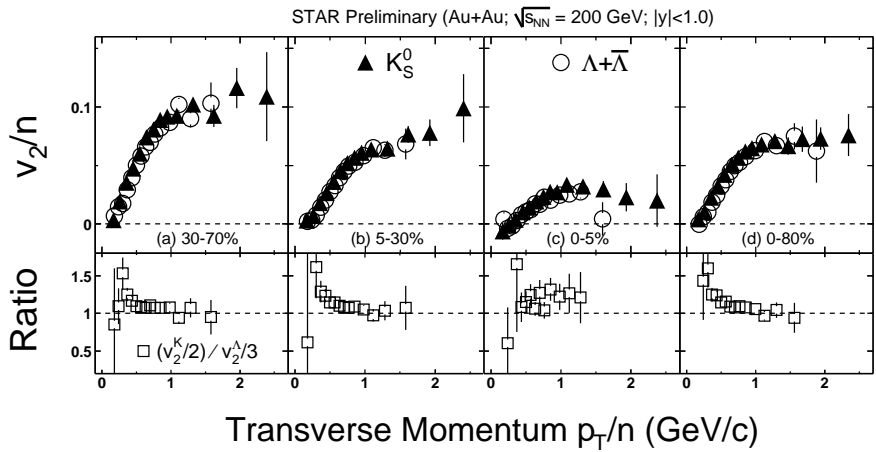}}
\caption{ The $v_2$ parameter for $K_{S}^{0}$ and $\Lambda +
\overline{\Lambda}$ scaled by the number of constituent quarks
($n$) and plotted verses $p_{T}/n$. } \label{fig3}
\end{figure}

The absence of a net suppression of $\Lambda+\overline{\Lambda}$
for $p_T$ from 1.8--3.5 GeV/c in central Au+Au collisions could
also indicate the presence of dynamics beyond the framework of
parton energy loss followed by fragmentation.
The stronger dependence on centrality (and thus parton density)
for baryon production indicated by the larger $R_{CP}$ would
naturally be expected from multi-parton mechanisms such as gluon
junctions~\cite{Vance}, quark coalescence~\cite{CoalVoloshinv2},
or recombination~\cite{CoalMullerRaa}.
Within the framework of these models, the measured $v_2$ and
$R_{CP}$ features may reflect the anisotropy and hadronization
properties of the bulk quark matter.
Fig.~\ref{fig3} shows $v_2$ of $K_{S}^{0}$ and $\Lambda +
\overline{\Lambda}$ as a function of $p_T$ where the $v_2$ and
$p_T$ values have been scaled by the number of constituent quarks
(n). Above $p_T$/n $\sim 0.8$~GeV/c, the $v_2$/n vs $p_T$/n is the
same, within errors, for both species.
In a scenario where hadrons at intermediate $p_T$ ($\sim
1-5$~GeV/c) are predominantly formed from bulk partonic matter by
quark coalescence, e.g. Ref.~\cite{CoalVoloshinv2}, $v_2/$n should
reveal the $v_2$ developed by partons prior to the hadronic phase.
The verification of this scenario, to the exclusion of other
possible explanations, would be strong evidence for the formation
of a quark-gluon plasma at RHIC.

The scenario discussed in Ref.~\cite{MesonBaryonGyulassy}, where
soft processes dominate the hyperon $v_2$ up to a higher $p_T$
than mesons, could also lead to a particle dependence for $R_{CP}$
that is qualitatively consistent with these measurements.
Quantitative calculations of $v_2$ and predictions for $R_{CP}$
from this scenario, however, are still needed. Up-coming
measurements of $R_{CP}$ for identified particles in d + Au
collisions at RHIC will also make it possible to study the effect
of initial state interactions on $R_{CP}$.

\section{Summary}
We have reported the measurement of $v_2$ for $p_T$ up to $\sim
6.0$~GeV/c for $K_S^0$ and $\Lambda + \overline{\Lambda}$ from Au
+ Au collisions at $\sqrt{s_{_{NN}}}=200$~GeV.
For $p_T < 1.2$~GeV/c, hydrodynamic model calculations agree well
with the $p_T$ and mass dependence of the measured $v_2$.
At this low momentum region $K_S^0$ and $\Lambda + \overline{\Lambda}$
$v_2$ lie on a single straight line when plotted versus $m_T-m_0$.
In the moderate $p_T$ region, however, the particle type and $p_T$
dependence of $v_2$ suggests hydrodynamics no longer describes the
collision dynamics.
The value of $v_2$ for $K_{S}^{0}$ saturates earlier and at a
lower value than the $\Lambda + \overline{\Lambda}$ $v_2$.
Measurements of $R_{CP}$ show that the suppression of particle
production in central collisions depends on particle-type in a
similar way.
The measurement of the particle-type and $p_T$ dependence of $v_2$
and $R_{CP}$ at moderate $p_T$ may provide a unique means to
establish the existence (and study the properties) of a
quark-gluon plasma that may be formed in collisions at RHIC.

\textbf{Acknowledgments:} This work has been partially supported
by NSF grant PHY-03-11859.

\end{document}